\documentstyle[iconip98,epsf]{article}
\newcommand{\mathcomma}{\, \, \mbox{}_{,}}
\newcommand{\mathperiod}{\, \, \mbox{}_{.}}

\begin{document}
\ninept

%
\title{Mean Field Approximation in Bayesian Variable Selection}

\author{Yukito Iba}
\email{iba@ism.ac.jp}

\institution{The Institute of Statistical Mathematics \\
106-8569, Minami-Azabu, Minatoku, Tokyo, Japan\\}
\maketitle

\begin{abstract}
Variable selection for a multiple regression model (Noisy Linear
Perceptron) is studied with a mean field approximation.
In our Bayesian framework,
variable selection is formulated as estimation of
discrete parameters that indicate a subset of the explanatory variables.
Then, a mean field approximation is introduced for the calculation of
the posterior averages over the discrete parameters.
An application to a real world example, Boston housing data, is shown.
\end{abstract}


\section{Mean Field Approximation}

Mean Field Approximation (MFA) is a well-known technique in
statistical physics. In general, MFA is defined as
a collection of techniques to calculate averages over multivariate
distributions, which uses the idea of ignoring
higher order correlations and solving self-consistent equations.
In this paper, we restrict ourselves to a type of MFA,
which treat distributions with discrete variables, specifically
binary variables.

MFA has a long history in physics.
It is, however, a recent topic in the area of
statistical information processing.
It is in the late 1980s that the mean field approximation
is used in the study of Boltzmann machine learning \cite{PH89}.
An application in image processing \cite{GG91} also
draws much attention of researchers in the area and works with
cluster versions of MFA appeared. 
Recently, a higher order version of MFA ( TAP equation )
is also applied to the decoding of error correcting codes \cite{KS98}
and Boltzmann machine learning.  A relation
between TAP equation and belief propagation in Bayesian network is 
discussed in \cite{KS98}.

The aim of this paper is to apply the idea of MFA to a
{\it variable selection } problem. In our Bayesian treatment,
variable selection for multiple regression is formulated as estimation of
discrete parameters that indicate a subset of the explanatory variables.
Then, a mean field approximation is introduced for the calculation of
the posterior marginals on the space of submodels, each of which is specified by a subset of the explanatory variables.


Although MFA is closely related to the Hopfield-Tank approach \cite{HT85} to combinatorial optimization, there is an important conceptual difference
\footnote{In a sense, the relation of MFA to the Hopfield-Tank method
is similar to that of MCMC to simulated annealing.
}. While the Hopfield-Tank method is a tool for optimization,
MFA is a method of approximation of the averages over a
distribution. MFA is a tool that is useful in the problems
where averages over a multivariate distribution play important roles. Two important examples of such problems are
the calculation of the averages over a Gibbs distribution in statistical
physics and the calculation of the posterior averages in Bayesian
statistics. In the present context, MFA gives posterior averages of discrete parameters, each of which gives the weight or ``the probability of the existence'' of the corresponding explanatory variable.
These averages take the values between 0 and 1.
At this point, our work is distinguished from other works on variable selection
with a Hopfield-Tank method, say, Sakai \cite{S93}.

\section{Generalization in Neural Networks}
\label{GNN}

Here we overview the issues on the generalization
in neural network models and present the motivation
of our work in the context of the interpolation between
ridge regression and variable selection.

Generalization from a finite set of examples is one of the
most important topics in the study of artificial neural network
in these ten years. Actually, it is a repeat of  a major
theme in statistics --- How can we beat overfitting and get a better
performance of prediction ?
For multiple regression models (noisy linear
perceptrons) and other types of feed-foward network models, two extreme approaches are known.

An extreme is
{\it ridge regression} or {\it linear weight decay},
where a penalty function of a quadratic form
of synaptic weights (regression coefficients) is added
to the target function that is optimized in the learning process \cite{M92}.
Information from given data is shared among a number of weights with this way of learning.
The other extreme is {\it variable selection} or {\it pruning },
where we select a subnetwork (a submodel)
with as small number of adjustable connections
as that is enough to represent the essence of the data.
Information is concentrated on small number of weights with the way of learning.

It is natural to interpolate these two extremes and develop
methods, such as  a ``fuzzy variable selection''.  A number of works
in this direction are classified into two categories.
A way to realize to such a interpolation is the use of
{\it generalized ridge regression}, or, {\it nonlinear weight decay}, where
a nonquadratic penalty function is used as the penalty to the magnitude
of synaptic weights (regression coefficients) \cite{IG85,WR91,NH92,W95,IS96,Neal96}.
Another approach to this problem is the introduction of weighted averages
over a set of models with different architectures (with different sets of explanatory variables), instead of using a single ``best'' model.
Methods of this category are most easily formulated
in a Bayesian framework, where the posterior distribution
on a space of submodels are treated as well as the posterior distribution
on a parameter space of each submodel \cite{MB88,IBA91,GM93,KM96,PS96}.

Our work belongs to the second category.
We are mostly interested in the algorithm for the
computation of posterior marginals in a space of submodels.
The introduction of the posterior distribution
on a set of submodels results in combinatorial explosion of
the requirement of computer resources, when
the submodels are not linearly ordered.
The calculation with exact enumeration is possible only
when the number of the variables is small \cite{MB88}.
A potentially powerful tool for this problem
is Markov Chain Monte Carlo algorithm (MCMC)
\cite{GM93,KM96,PS96,IBA91}.
The Monte Carlo approach is, however,
still computationally expensive and often suffers from slow convergence.

At this point we introduce a Mean Field Approximation.
As already suggested in the previous section,
the essence of our idea is the application MFA
for calculation of the posterior average on the space of submodels.
In the next section, we will give  a formulation and basic notations
for the implementation of the idea to a linear network, i.e., a multiple regression model.

\section{Variable Selection for a Multiple Regression Model}
\label{VSMR}

The likelihood function of a multiple regression model is
\begin{equation}
-E_L= -\frac{1}{2\sigma^2}\sum_{\alpha=1}^N (y^\alpha -
\sum_{i=1}^M a_i x_i^\alpha)^2 \mathcomma
\end{equation}
\begin{equation}
L(\{y^\alpha\}|\{a_i\})= \frac{1}{(2\pi \sigma^2 )^{N/2}} \exp(-E_L) \mathcomma
\end{equation}
where $y^\alpha$  and $\{x_i\}^\alpha $  are the $\alpha$th observation of the response variable (the output of the network)  and  the explanatory variables (the inputs of the network) respectably.
$N$ is the number of the observations and $M$
is the maximum number of explanatory variables available in data.
We introduce a set of indicators $\{\widetilde{S}_i\},  \, \widetilde{S}_i \in \{0,1\}$ for the description of submodels.
If and only if $\widetilde{S}_i=1$,  the variable $x_i$ appears in the submodel specified by $\{\widetilde{S}_i\}$.
Using the indicators $\{\widetilde{S}_i\}$, a submodel
 is written as
\begin{equation}
\label{like1}
-\widetilde{E}_L= -\frac{1}{2\sigma^2}\sum_{\alpha=1}^N (y^\alpha
- \sum_{i=1}^M a_i x_i^\alpha \widetilde{S}_i)^2 \mathcomma
\end{equation}
\begin{equation}
\widetilde{L}(\{y_\alpha\}|\{a_i\},\{\widetilde{S}_i\})=
\frac{1}{(2\pi \sigma^2 )^{N/2}} \exp(-\widetilde{E}_L) \mathperiod
\end{equation}
With this submodel, the maximum likelihood estimator of
the regression coefficient $a_i$ is
obtained as the solution of the corresponding normal equation
\begin{equation}
\label{normal}
V_{x_ix_i}a_i + \sum_j V_{x_ix_j}a_j \widetilde{S}_j = V_{x_iy}
\mathcomma
\end{equation}
when $\widetilde{S}_i=1$.
Here, sufficient statistics $V_{y^2}, V_{x_iy}, V_{x_ix_j}$ are defined as
$V_{y^2}  =  \sum_\alpha  (y^\alpha)^2$,
$V_{x_iy}  =  \sum_\alpha  (x_i^\alpha y^\alpha)^2$,
$V_{x_ix_j}  =  \sum_\alpha  (x_i^\alpha x_j^\alpha)^2$.

When we define $S_i = 2 \widetilde{S}_i -1$  and rearrange the terms,
we have
\begin{equation}
\label{like3}
-\widetilde{E}_L= \frac{1}{2\sigma^2}
(C + \sum_i H_i S_i + \sum_i \sum_{j>i} J_{ij} S_i S_j) \mathcomma
\end{equation}
\begin{equation}
\label{CC}
C = - V_{y^2} + \sum_i a_i V_{x_iy}
- \frac{1}{2} \sum_i \sum_{j>i} a_i a_j V_{x_i x_j}
- \frac{1}{2} \sum_i a_i^2 V_{x_ix_i} \mathcomma
\end{equation}
\begin{equation}
\label{HH}
H_i = a_i V_{x_iy} - \frac{1}{2} \sum_{j \neq i} a_i a_j V_{x_i x_j}
-\frac{1}{2} a_i^2 V_{x_ix_i} \mathcomma
\end{equation}
\begin{equation}
\label{JJ}
J_{ij} = - \frac{1}{2} a_i a_j V_{x_i x_j}
\end{equation}
using $\widetilde{S}_i^2= \widetilde{S}_i$.
For later convenience, the indicator $S_i$ that takes
the values of  $\pm 1$ is introduced.

To define a Bayesian model, we should specify prior distributions.
The prior distribution of  the indicators $\{S_i\}$ is assumed as
\begin{equation}
\pi(\{S_i\})=
\frac{\exp(-h \sum_i S_i)}{(\exp(h)+\exp(-h))^M} \mathcomma
\end{equation}
where the hyperparameter $h$ determines the degree of the penalty to
submodels with larger number of explanatory variables.
In this paper, we also assume that the prior
of the coefficients $\{a_i\}$ is the uniform density
in $\mbox{\bf R}^M$.
Then the posterior distribution $\widetilde{P}_{pos}(\{a_i\},\{S_i\})$
of $\{S_i\}, \{a_i\}$ is formally defined as a Gibbs distribution
\begin{equation}
\widetilde{P}_{pos}(\{a_i\},\{S_i\})=\frac{\exp(-\widetilde{E}_{pos})}{\widetilde{Z}_{pos}}
\end{equation}
with the energy
$ -\widetilde{E}_{pos}=
-\widetilde{E}_L - h \sum_i S_i $ and the normalization
$\widetilde{Z}_{pos}=\sum_{\{S_i\}} \int \Pi_i d a_i
\exp(-\widetilde{E}_{pos})$
\footnote{The integral in the definition of $\widetilde{Z}_{pos}$
diverges with the uniform prior for $\{a_i\}$.
Apparently, this is not significant in our approximation,
where a point estimate $\{a^*_i\}$ of the regression coefficients is used.
It might cause, however, some difficulties, e.g.
in a Bayesian interpretation of the strength $h$ of the penalty.}.
Here and hereafter, $\sum_{\{S_i\} }$ denote the summation over
$2^M$ configurations of  $\{S_i\}$.

\section{MFA for Bayesian Variable Selection}
\label{MFABVS}

Now, we discuss an application of Mean Field Approximation
to variable selection for multiple regression.
First, we consider the posterior distribution of $\{S_i\}$ conditioned with a given set $\{a_i\}$ of regression coefficients,
\begin{equation}
\label{postS}
P_S(\{S_i\}|\{a_i\})=
\frac{\exp(-\widetilde{E}_{pos})}{Z_S} \mathperiod
\end{equation}
The normalization constant $Z_S$ is defined by
\begin{equation}
\label{Zs}
Z_S= \sum_{\{S_i\} } \exp(-\widetilde{E}_{pos}) \mathperiod
\end{equation}
Consider a distribution
\begin{equation}
\label{MFdist}
Q(\{S_i\}) = \Pi_i \frac{1 + m_i \cdot S_i}{2}
\end{equation}
and the KL-divergence
\begin{equation}
D(Q; P_S)= \sum_{\{S_i\}} Q(\{S_i\}) \log \frac{Q(\{S_i\})}{P_S(\{S_i\}) | \{a_i\})}
\end{equation}
between $Q$ and $P_S$. An easy calculation shows that
\begin{eqnarray}
D(Q; P_S)= & - \frac{1}{2\sigma^2} ( C+ \sum_i \sum_{j>i} J_{ij} m_im_j
+ \sum_i H_i m_i ) \nonumber \\
 & + h \sum_i m_i  + \log Z_S \nonumber \\
          & + \sum_i  \left \{ \frac{1+m_i}{2} \log \frac{1+m_i}{2} +
                              \frac{1-m_i}{2} \log \frac{1-m_i}{2}  \right \}_.
\end{eqnarray}
The values $\{m_i^*\}$ of $\{m_i\}$ that minimize
$D(Q; P_S)$ is a solution of  the ``self-consistent equation''
\begin{equation}
\label{self}
m_i^* = \tanh \left( \frac{\sum_{j \neq i} J_{ij}m_j^* + H_i}
{2\sigma^2}  -h \right) \mathperiod
\end{equation}

From the inequality $D(Q; P_S) \geq 0$, the relation
$\log Z_S  \geq  \log Z^*_{MF}$
is derived, where
\begin{eqnarray}
\label{MFfree}
\lefteqn{\log Z^*_{MF}= \nonumber} \\
 & \frac{1}{2\sigma^2}
(C + \sum_i H_i m^*_i + \sum_i \sum_{j>i} J_{ij} m^*_i m^*_j) - h \sum_i m^*_i
\nonumber \\ &
-\sum_i \left \{ \frac{1+m^*_i}{2}\log \frac{1+m^*_i}{2}
+ \frac{1-m^*_i}{2}\log \frac{1-m^*_i}{2} \right \} \mathperiod
\end{eqnarray}
The equality is hold only when the distributions
$Q$ and $P_S$ are identical.
The form of the distribution (\ref{MFdist}) suggests that
the $i$th component $m^*_i$ of a solution of the equation (\ref{self})  is
regarded as an approximation of the averages of the indicator  $S_i$ over the distribution (\ref{postS}). Thus the posterior probability $p_i$ that the variable $x_i$ is contained in a submodel is approximated by
\begin{equation}
\label{MFprob}
p^*_i = \frac{1+m^*_i}{2} \mathperiod
\end{equation}
We can also regard $\log Z^*_{MF}$
as an approximation of $\log Z_S$. The self-consistent equation (\ref{self}) and the expressions
(\ref{MFprob}) and (\ref{MFfree}) are
the essence of the MFA in our problem.

So far, we are working with a given set of regression coefficients $\{a_i\}$ .
In our problem, they are unknown parameters.
A way to deal with them is to replace them with point estimates that
maximize the likelihood
\begin{equation}
l(\{a_i\},\sigma^2)=
\log \sum_{\{S_i\}} \widetilde{L}(\{y^\alpha\}|\{a_i\},\{S_i\})
\pi(\{S_i\})
\end{equation}
marginalized over $\{S_i\}$
\footnote{
Although the use of a point estimation of $\{a_i\}$ is
not fully justified in a Bayesian paradigm and leads to an approximation
beyond the conventional use of MFA,
it greatly simplified the situation.}.
It is represented by the normalization constant
$Z_S$ of (\ref{Zs}) as
\begin{equation}
\label{ml}
l(\{a_i\},\sigma^2)=
\log Z_S
- M \log (e^h+e^{-h})  - \frac{N}{2} \log {2\pi\sigma^2} \mathperiod
\end{equation}
This marginal likelihood (\ref{ml}) is approximated by the MFA as
\begin{equation}
\label{mff}
l_{MF}(\{a_i\}, \sigma^2)= \log Z^*_{MF}
- M \log (e^h+e^{-h})  - \frac{N}{2} \log {2\pi\sigma^2} \mathperiod
\end{equation}
The values $\{a_i^*\}$ that maximize (\ref{mff}) is
a solution of a ``soft version'' of the normal equation
\footnote{
Note that the values $\{m^*_i\}$ should be considered as
a function of $\{a_i^*\}$ in the derivation.
However, with the use of $\frac{\partial D(Q; P_S)} {\partial m_i} |_{\{m_i\}=\{m_i^*\}} =0$,
the result coincides with that of a naive calculation.}
\begin{equation}
\label{mfnormal}
V_{x_ix_i}a_i^* + \sum_{j \neq i} V_{x_ix_j}a_j^* p^*_j = V_{x_iy} \mathperiod
\end{equation}
The equation (\ref{mfnormal}) is, in fact, a equation
nonlinear in $\{a_i^*\}$, because the probability $\{p^*_j\}$ is a function
of $\{a^*_i\}$ implicitly determined with (\ref{MFprob}) and (\ref{self},\ref{HH},\ref{JJ}).
In actual implementation, the following algorithm is used to
solve the set of equations.

(1) Set the initial values $\{m^0_i \}$ of $\{m_i\}$ and $t: =0$.

(2) Solve the equation of $\{a_i^t\}$
\begin{equation}
\label{mfnormal2}
V_{x_ix_i}a_i^t + \sum_{j \neq i} V_{x_ix_j}a_j^t p^t_j = V_{x_iy} 
\end{equation}
with $p^t = (1+m_i^t)/2$.

(3) Iterate the equation
\begin{equation}
m_i^{t+1} := (1-\delta) \cdot m_i^{t} +
\delta \cdot \tanh  \left(  \frac{\sum_{j \neq i} J_{ij}m_j^{t} + H_i}
{2\sigma^2} -h  \right) \mathperiod
\end{equation}
Here, the variables $\{H_i\}$,$\{J_{ij}\}$
are defined by (\ref{HH},\ref{JJ}) with $a_i=a^t_i$.
$\delta$ is a constant that satisfies $0 < \delta \leq 1$.

(4) Set $t: =t+1$. Check the convergence and return to the step (2), if necessary. If convergence is ensured,
set $p_i^*:=(1+m_i^t)/2$ and $a^*_i:=a^t_i$ and exit.

We can also estimate the variance $\sigma^2$ of the residual with the maximization of (\ref{mff}). For this purpose, we add a step for the estimation of $\sigma^2$ to the iteration. In our implementation,
the value of $\sigma^2$ is updated once per the iteration.

\section{An Example}
\label{AE}

In this section, we discuss an application of the proposed 
method to a real world
example. We use the dataset taken
from p.244 of Belsley et al. \cite{BKW89}.
This dataset, known as Boston housing data, is also analyzed by other authors with different methods~\cite{BKW89,Neal96,Shimo93}.
There are thirteen explanatory variables ($M=13$) and one response variable, the logarithm of the median value of owner-occupied homes for each area of Boston. The sample size is $N=506$.

Our results on this data are shown in Fig.1 and Fig.2.
In the experiment, the variance $\sigma^2$ is estimated from the data.
Although we observe local optima with some parameters of $h$,
they are not serious and the solution with
the highest frequency at each value of $h$ is plotted in the figures.
\begin{center}
\begin{minipage}{90pt}
\leavevmode
\epsfxsize=90pt
\epsfysize=95pt
\epsfbox{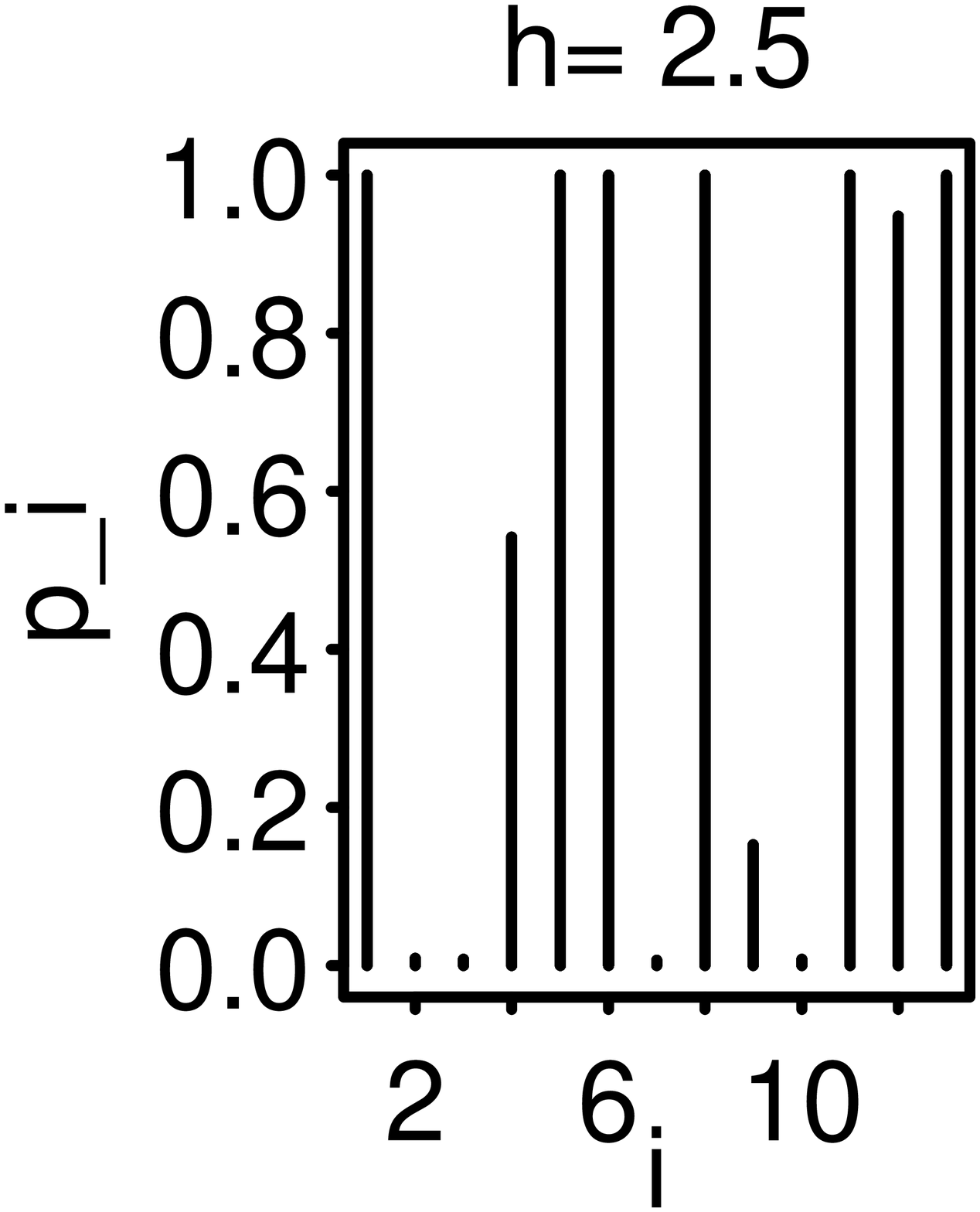}
\end{minipage}
\begin{minipage}{90pt}
\leavevmode
\epsfxsize=90pt
\epsfysize=95pt
\epsfbox{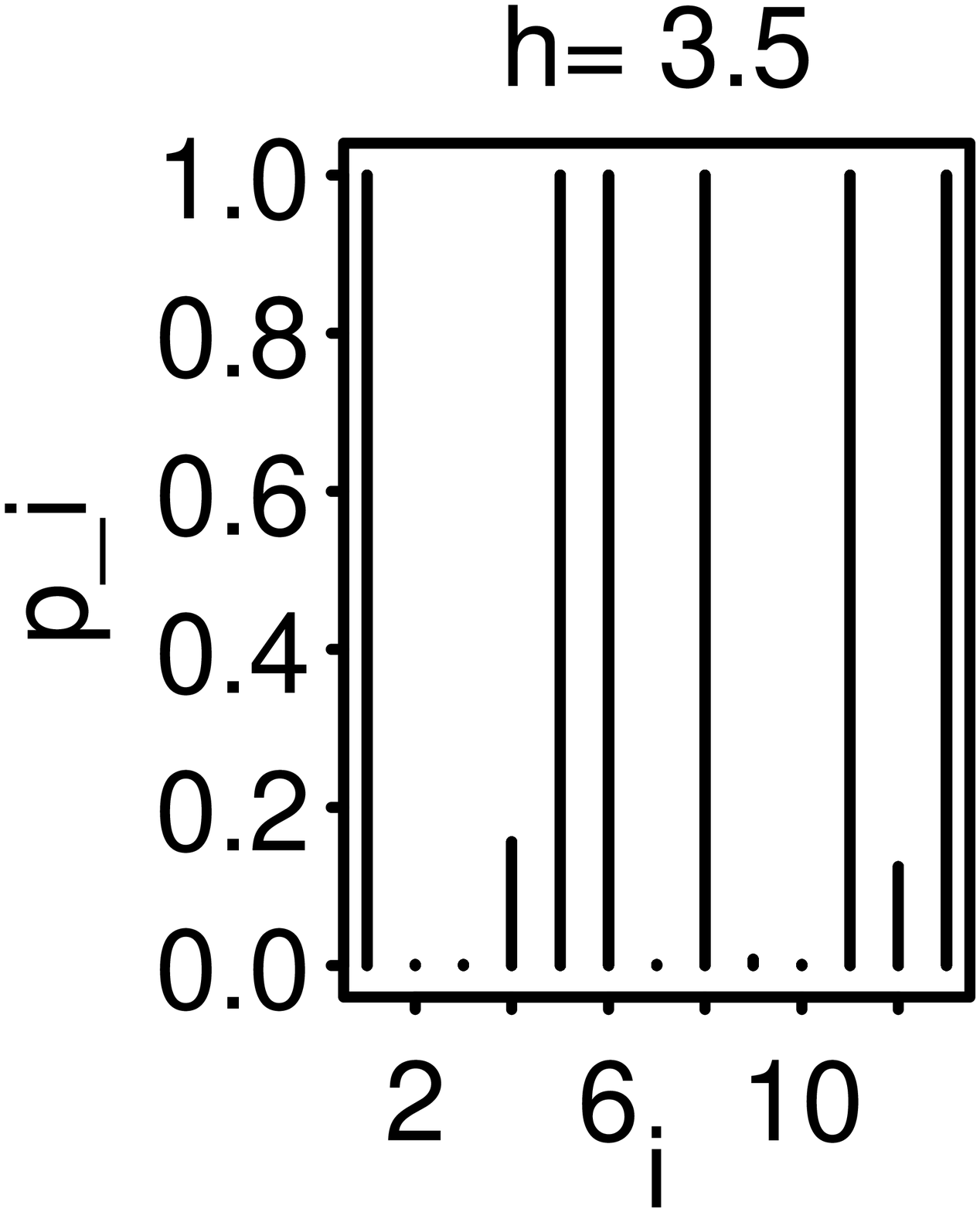}
\end{minipage}

\begin{minipage}{90pt}
\leavevmode
\epsfxsize=90pt
\epsfysize=95pt
\epsfbox{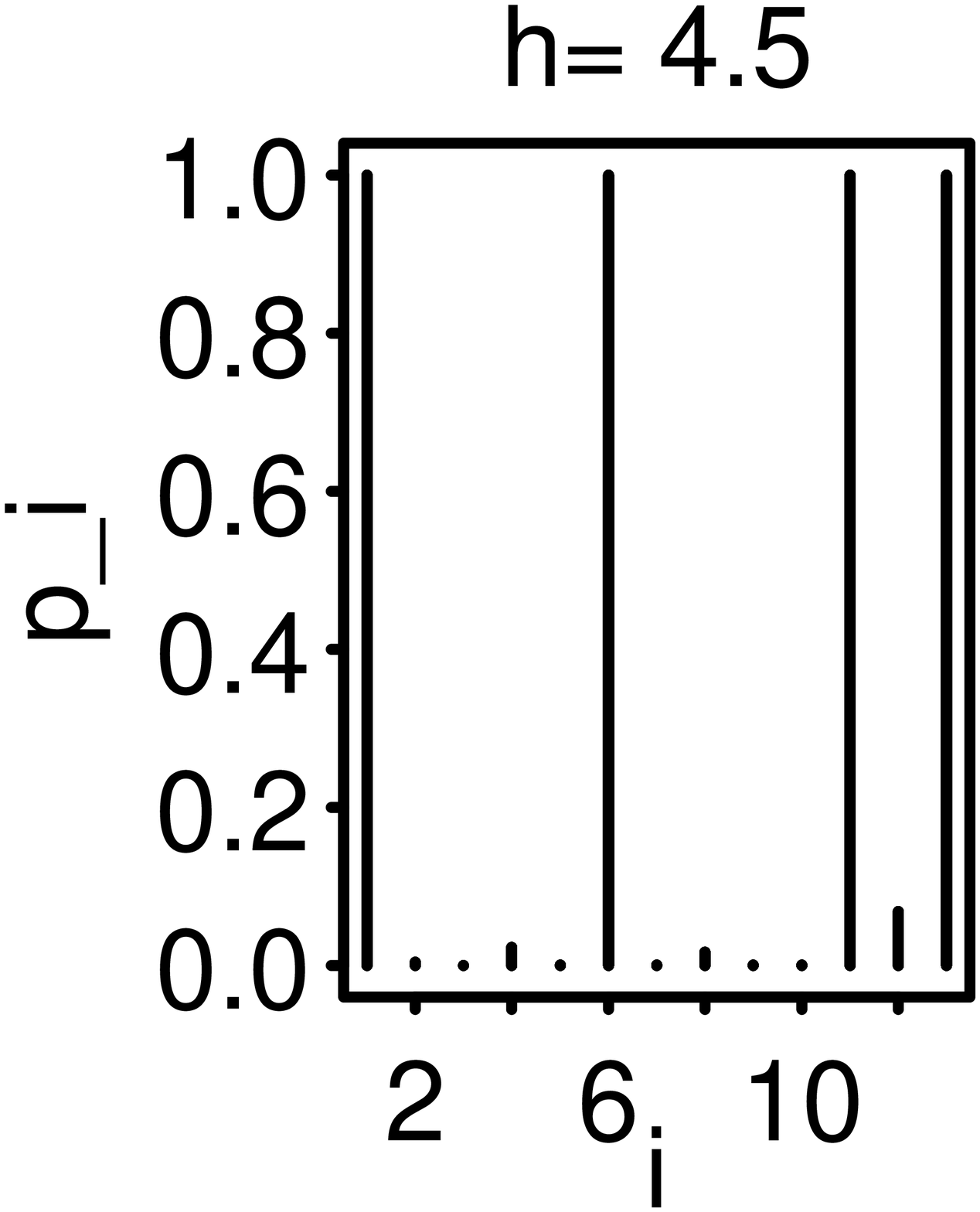}
\end{minipage}
\begin{minipage}{90pt}
\leavevmode
\epsfxsize=90pt
\epsfysize=95pt
\epsfbox{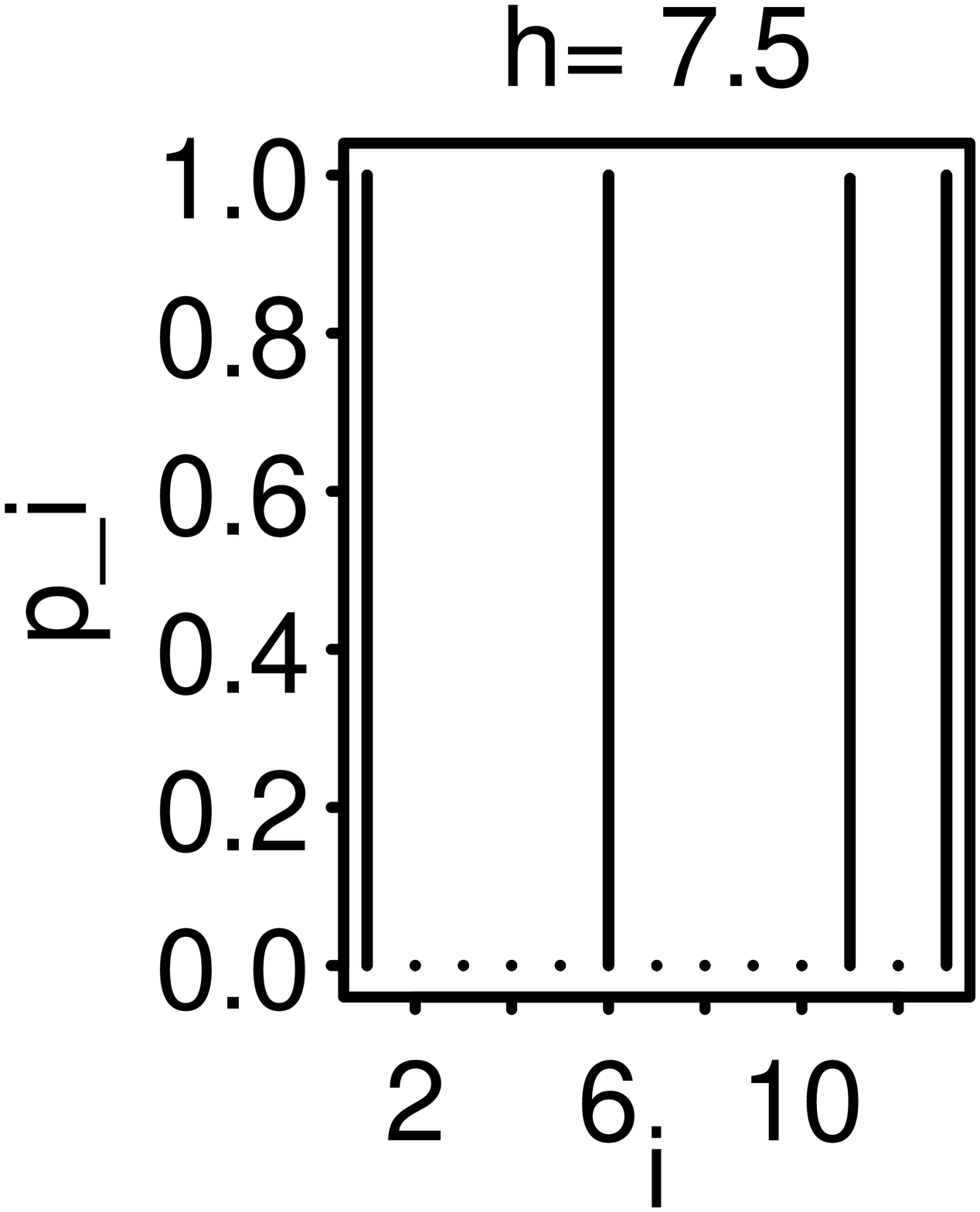}
\end{minipage}
\end{center}
Fig.1: \ The results with some typical
values of $h$. The index $i$ of a variable is shown in the horizontal axis
and the probability $p^*_i=(1+m_i^*)/2$
is shown in the vertical axis.\\
\begin{center}
\begin{minipage}{210pt}
\leavevmode
\epsfxsize=210pt
\epsfysize=220pt
\epsfbox{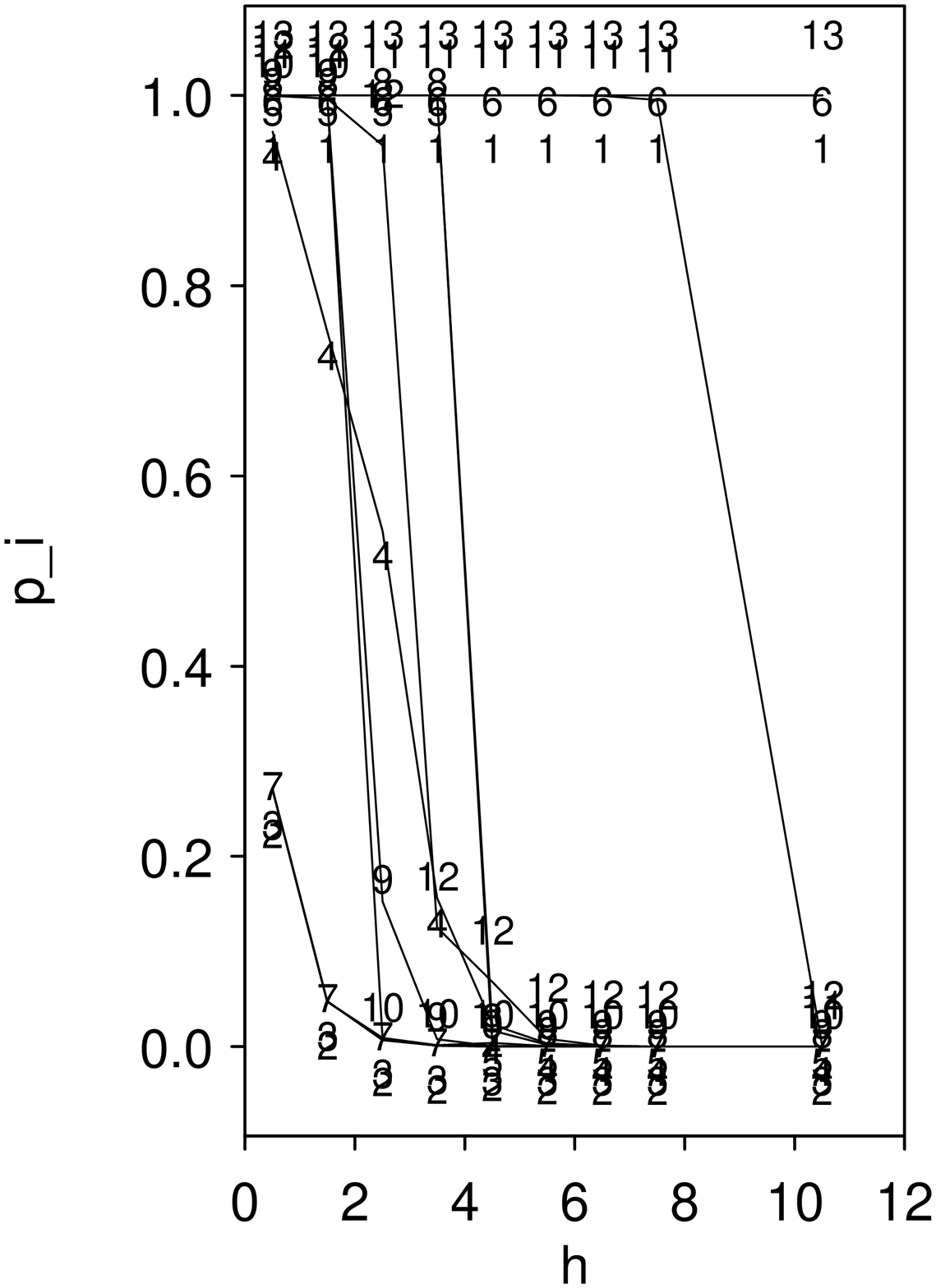}
\end{minipage}
\end{center}
Fig.2: \ The change of $p^*_i$ is shown with $h$.
The number used as a symbol indicates the index~$i$ of the variable. 
The positions of the symbols are slightly perturbed
by noise to prevent complete overlap between them.
Lines are accurate.

In Fig.2, the probability $p^*_i$ takes the value 
$p^*_i \sim 1$, for $i=1,6,13$, even with a
larger value of the penalty $h$. It seems consistent with the result
in \cite{Shimo93}. For most of the other variables, 
the probability $p^*_i$  shows a sudden jump between 0 and 1, or, a simple 
decay with $p^*_i \sim \frac{\exp(-h)}{\exp(h)+\exp(-h)}$, as the value of
$h$ is increased.
However, for some $i$, say $i=4$, it shows a gradual decay
and has nontrivial values between $0$ and $1$ in a range of $h$.

\section{Future Problems}
\label{FP}

In the present treatment, the strength $h$ of the penalty to the number
of variables is a free hyperparameter. It is not a serious fault
when we use the proposed method as a tool for {\it regression diagnostics },
i.e., we observe the behavior of solutions in different values of $h$ and get information on data.  A criterion to determine the optimal value of $h$ is, however, required, when the method is used as
a fully automatic tool for the prediction. Although there are a few possible
approaches, their implementations and tests are left
for future studies.

Another important problem is the study of the multiple
solutions in our iterative procedure. For example,
when near loss of the linear independence among explanatory variables ({\it colinearity}) exists, the algorithm is likely to have multiple fixed points.
Note that the existence of
local optima is usually regarded as a mere difficulty of algorithms
in information processing.
There seems, however, often a possibility of exploration
of the nature of the data from the emergence of multiple extremum.
It is interesting if the proposed 
algorithm provides an example for such a study. 
For this purpose, it requires more investigation into
the behavior of the algorithm in complex situations.

We acknowledge Dr.~Kabashima for fruitful discussions on the work.
This work is supported by RWC Project
of Ministry of International Trade and Industry.

\end{document}